\documentclass[a4paper]{jpconf}
\usepackage{graphicx}
\usepackage{natbib}

\begin{document}
\title{Location and origin of gamma-rays in blazars}

\author{B. Rani, T. P. Krichbaum, J. A. Hodgson, J. A. Zensus}
\address{Max-Planck-Institut f{\"u}r Radioastronomie (MPIfR), Auf dem H{\"u}gel 69, D-53121 Bonn, Germany}
\ead{brani@mpifr-bonn.mpg.de}

\author{On behalf of the {\it Fermi}-LAT Collaboration}

\begin{abstract}
One of the most intriguing and challenging quests of current astrophysics is to understand the physical conditions 
and processes responsible for production of high-energy particles, and emission of $\gamma$-rays. A combination of 
high-resolution Very Long Baseline Interferometry (VLBI) images with broadband flux variability measurements is a 
unique way to probe the emission mechanisms at the bases of jets. Our analysis of $\gamma$-ray flux variability observed 
by the {\it Fermi}-LAT (Large Area Telescope) along with the parsec-scale jet kinematics suggests that the $\gamma$-ray emission 
in blazar S5 0716+714 has a significant correlation with the mm-VLBI core flux and the orientation of jet outflow on parsec 
scales. These results indicate that the inner jet morphology has a tight connection with the observed $\gamma$-ray flares. 
An overview of our current understanding on high-energy radiation processes, their origin, and location is presented here. 
\end{abstract}

\section{Introduction}
The origin of high-energy emission has long been a key question in Active Galactic Nuclei (AGN)
physics. While they represent only a minority of AGN, blazars constitute a unique laboratory to
probe the acceleration processes in relativistic jets emanating from the central engine.
Since its launch in 2008, the {\it Fermi}-LAT (Large Area Telescope) has revolutionized
our knowledge of the $\gamma$-ray sky with a combination of high sensitivity, wide
field-of-view, and a large energy range (about 20 MeV to more than 300 GeV). Its nominal
sky-survey operating mode has enabled a continuous monitoring of the complete $\gamma$-ray
sky at an unprecedented level, making remarkable discoveries. The third {\it Fermi}-LAT
catalog \citep[3FGL,][]{acero2015}, based on the first 48 months of data, consists of more than 3000
$\gamma$-ray  sources including $\sim$60$\%$ AGN - most of which are blazars \citep[$\sim$98$\%$][]{ackermann2015}. 
Blazars therefore form a major population of the $\gamma$-ray sky and are potential targets 
for understanding the physical processes responsible for production of high-energy emission. 
Moreover, blazars also account for the majority of the high energy $\gamma$-ray background 
above 100~GeV \citep{ajello2015}.

BL Lacertae objects (BL~Lacs) and flat spectrum radio quasars (FSRQs) are clubbed together and called blazars.
In spite of the dissimilarity of their optical spectra -- FSRQs show strong broad emission lines, while
BL~Lacs have only weak or no emission lines in their optical spectra -- they share the same peculiar continuum
properties (strong variability and polarization properties). They are characterized by powerful non-thermal
emission ranging from radio to the $\gamma$-ray bands and exhibit strong variability, over a variety
of time scales from minutes to months and often the radio components detected in VLBI observations
exhibit superluminal motion. These properties are interpreted as resulting from the emission of
high-energy particles accelerated within a relativistic jet aligned with the direction of sight.
In many cases, the bulk of the observed flux emerges in $\gamma$-ray photons in GeV (TeV for some sources) 
energy range populating a major fraction of the high-energy sky \citep{ackermann2015, acero2015}. One of the most 
intriguing and challenging quests of current astrophysics is to understand the physical conditions and processes 
responsible for the  production of high-energy particles, and the emission of $\gamma$-rays.

\section{Particles accelerated in relativistic jets} 

Detailed investigation of multi-wavelength flux and spectral variability of individual sources including
cross-band (radio, optical, X-ray, $\gamma$-ray, and polarization), relative timing analyses of
outbursts, and/or VLBI component ejection/kinematics suggest that $\gamma$-rays are associated with
the compact regions of relativistic jets energized by the central SMBHs \citep[e.g.,][and references therein]{agudo2011, jorstad2001, jorstad2010, marscher2010, schinzel2012, rani2013a, rani2015}. For 
individual objects, the close association of the $\gamma$-ray flare 
with a continuous change of the optical polarization angle provides evidence for the presence of highly ordered 
magnetic fields in the regions of $\gamma$-ray production \citep{abdo2010c, marscher2008}. 
Recent studies (at least for some blazars) suggest a
significant correlation between the variations of $\gamma$-ray flux and the direction of the jet outflow
in the sub-parsec scale region \citep{rani2014}, which implies that the observed inner jet morphology has
a strong connection with the observed $\gamma$-ray flares.  However, the fine details of acceleration
mechanisms and radiation processes are still missing.

Blazars exhibit a characteristic bimodal spectral energy distribution (SED). The SED low-energy
peak (from radio to optical/UV and X-rays for some sources) is commonly interpreted as synchrotron emission
from high-energy electrons, while no consensus on the origin of the high-energy peak (from X-rays to
$\gamma$-rays) has been reached. In the framework of leptonic models, this second peak corresponds
to photons produced via inverse-Compton scattering of soft seed photons by the same electrons, with
these seed photons originating either from within the jet (Synchrotron Self-Compton, SSC) or from regions
external to the jet, such as the accretion disk, the dust torus or the broad emission line clouds
(External Compton, EC). In hadronic models, the high energy peak is due to photons produced in
photo-hadron interactions or by proton or charged $\mu$/$\pi$ synchrotron emission and subsequent
cascades \citep[e.g.,][]{bottcher2007, bottcher2012, bottcher2013}.
Concerning the high-energy emission, leptonic models are more favored in the literature;
however, hadronic processes remain a viable option.

No clear consensus has been achieved on the location of emission regions. The radio--$\gamma$-ray 
correlations \citep{fuhrmann2014, marscher2008, marscher2010, rani2013a}  suggests the latter 
leading the former, which puts the location of high-energy emission regions closer to the central 
black hole.   The observed $\gamma$-ray spectral breaks at a few GeVs require the emitting region to 
be located within the BLR if the photon-photon pair production scenarios are responsible for their 
origin \citep{abdo2009, finke2010, rani2013b, rani2013_3c273, poutanen2010}. On the other hand, coincidence 
of $\gamma$-ray flares with the appearance of new components from the base of the jet 
indicates distances longer than few parsecs \citep{jorstad2013, agudo2011, schinzel2012}. 
Interaction of moving features with the stationary features can amplify the magnetic field and
accelerate particles \citep{marscher2014}.  Coincidence of $\gamma$-ray flares with the passage of moving components through 
stationary features in jets has been observed in several sources \citep{schinzel2012, rani2015}, which implies 
that $\gamma$-ray flares are produced downstream of the core at a distance of more than a few parsecs.

Particle acceleration in relativistic jets is also an open question. Relativistic shocks seem to be 
effective in low magnetization regimes; however in the case of high magnetization, shocks are weak and poor 
accelerators. Particle acceleration through second order stochastic processes is rather slow in 
highly magnetized flows; but it is still a viable option \citep{blanford1973}. In highly magnetized environments, 
particle acceleration via magnetic reconnection is very likely to occur \citep{kagan2014}. Most likely 
all of these processes are in action on different scales in jets. Closer to the central engines, magnetic 
reconnection could inject energetic particles in compact regions to produce rapid $\gamma$-ray flares.  
Further downstream the jet  in low magnetized environments, the observed $\gamma$-ray variations 
could be a natural consequence of relativistic shocks.  Magnetoluminescence and/or electromagnetic detonation 
is lately proposed to explain the fastest observed TeV flares.  The basic idea is that the tightly
wound magnetic field in the inner jet regions is subject to dynamical instability which could be responsible 
for the rapid $\gamma$-ray variability. In other words, impulsive particle acceleration can take place while 
the magnetic flux ropes untangle \citep{blanford2015}, which could be responsible for high-energy flares.

 \begin{figure}
\begin{minipage}{0.5\linewidth}
 \includegraphics[scale=0.45,angle=0, trim=130 0 120 2, clip]{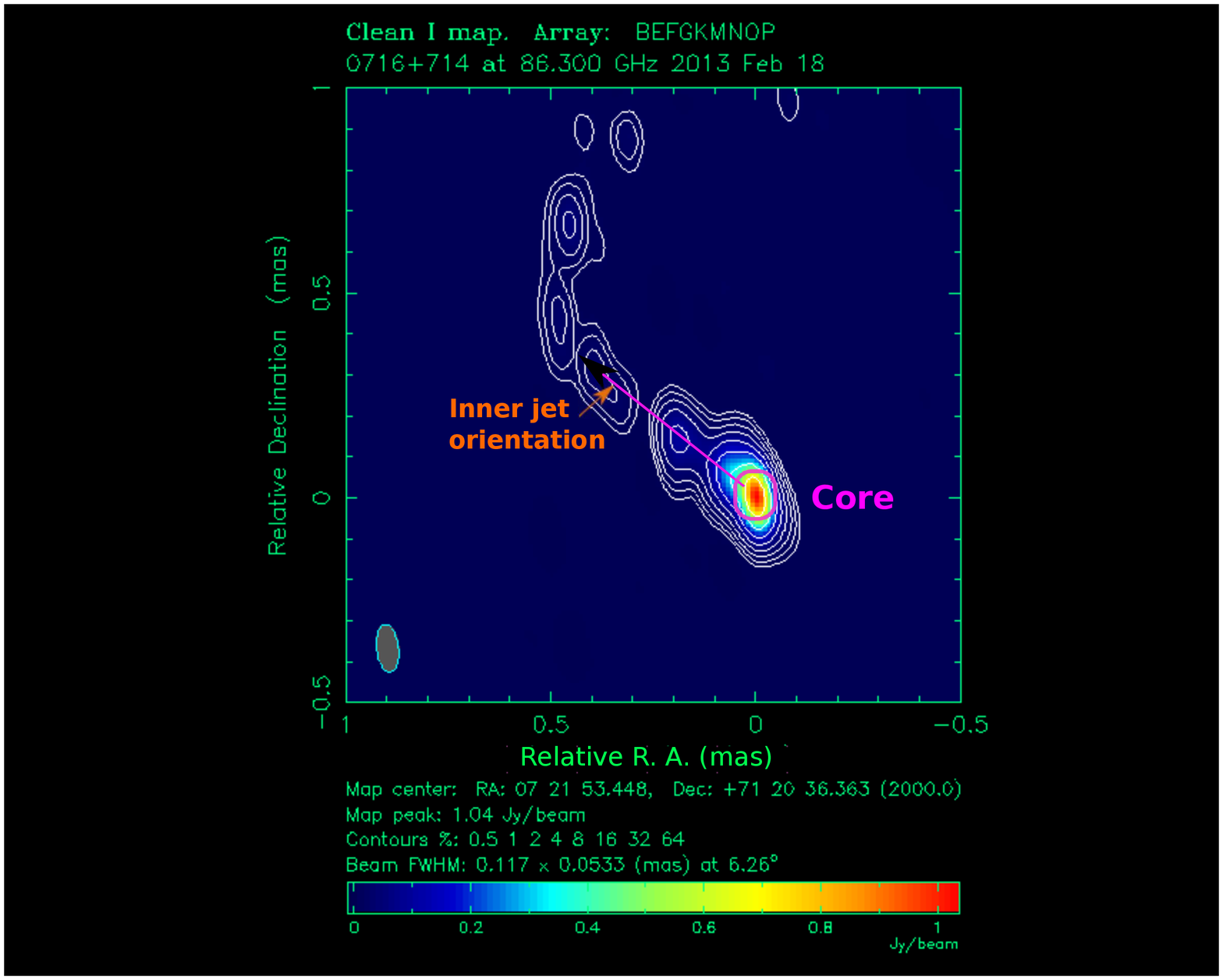}
\end{minipage}\hfill
\begin{minipage}{0.5\linewidth}
\includegraphics[scale=0.323,angle=0,trim=0 0 0 2, clip]{PA_var.eps}
\includegraphics[scale=0.323,angle=0,trim=0 0 1.8 2, clip]{core_flx_var.eps}
\includegraphics[scale=0.323,angle=0,trim=5 0 0 2, clip]{gamma_var.eps}
\end{minipage}\hfill
\caption{Left: The 86~GHz VLBI image of S5 0716+714 observed on February 18, 2013. Right: (a) 
Orientation variations in inner region ($\leq$0.2~mas) of the jet, (b) the 7~mm VLBI core flux density 
variations,  and (c) the monthly averaged $\gamma$-ray photon flux light curve.     }
\label{fig1}
\end{figure}

\section{Causal connection between gamma-ray emission and jet morphology}
Our recent studies on the BL Lac S5 0716+714 suggest a significant correlation between the
variations of $\gamma$-ray flux and the direction of the jet outflow in sub-parsec scale region 
\citep[see][for details]{rani2014}. The core flux density and inner-jet orientation (position angle) variations 
were investigated using the 7~millimeter Very Long Baseline Interferometry (VLBI) observations of 
S5 0716+714 for a time period between August 2008 and September 2013. In Fig.\ \ref{fig1} (left), 
we show an example map of the source. The variations 
in the core flux density and orientation of the sub-parsec scale jet i.e.\ position angle (PA) are 
shown in Fig.\ \ref{fig1} (a and b). 
The $\gamma$-ray flux variations (Fig.\ \ref{fig1} c) over the same time period were examined using the 
observations obtained  by the {\it Fermi}-LAT (Large Area Telescope).  

\begin{figure}
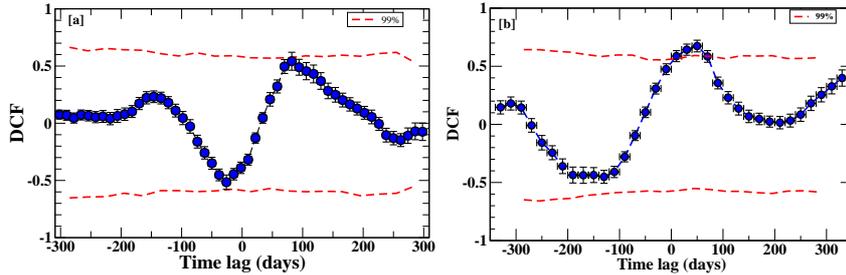

\begin{minipage}{0.7\linewidth}
\includegraphics[scale=0.22,angle=0,trim=0 0 0 0, clip]{dcf_gamma_core.eps}
\includegraphics[scale=0.22,angle=0,trim=0 0 0 0, clip]{dcf_gamma_pa.eps}
\end{minipage}
\begin{minipage}{0.3\linewidth}
  \caption{Cross-correlation analysis curves: $\gamma$-ray vs.\ core flux density (a) and $\gamma$-ray vs.\ position 
angle (b). The dashed lines show the 99$\%$ confidence levels.    }
\end{minipage}
\label{fig2}
    \end{figure}

The discrete cross-correlation function \citep[DCF,][]{edelson1988} method was used to investigate the correlation between 
the observed $\gamma$-ray and jet morphology (core flux density and PA) variations. The DCF analysis 
results are presented in Fig.\ \ref{fig2}. The analysis suggests that the high-energy radiation has 
a tight correlation between the VLBI core flux density and the direction of the jet flow flow (Fig.\ \ref{fig2}). 
We observed a time lag of 82$\pm$39 days between$\gamma$-ray and core flux density variations, which 
puts the location of $\gamma$-ray emission region upstream of the 7~mm VLBI core by 3.8$\pm$1.9~pc 
(de-projected). Given the location of the 7~mm core at a distance of $\sim$6.5~pc \citep{rani2015}, 
we conclude that the $\gamma$-ray emission region is at a distance of 2.7$\pm$1.9~pc from the central engine. The 
correlated variations observed in the innermost PA and the high-energy flux of 
S5 0716+714 add a unique new point to constrain the location and origin of high-energy emission in 
blazars.

\section{Future Perspectives}
High-frequency VLBI is a promising technique to pinpoint the high-energy emission sites.  
With the current GMVA (Global mm-VLBI Array) resolution at 86~GHz frequency ($\sim$50~$\mu$as), one could 
scale down to less than a thousand Schwarzschild radii. The Event Horizon Telescope (EHT) will offer an 
angular resolution of $\sim$20~$\mu$as, which means that the future of high-resolution VLBI is very bright. 
Participation of ALMA will bring a revolutionary leap in capabilities for mm-VLBI. The {\it Fermi} mission with its 
extraordinary capabilities will keep observing the $\gamma$-ray sky at least for the next few years. 
In the upcoming years, TeV astronomy will also grow making several more new and major discoveries. 
Not to mention, high-energy polarization missions (Astro-H, GEMS, X-calibur, PoGOLite, Polar, Harpo, 
and many more) are also on their way to teach us more about high-energy emission models.

\ack
The $Fermi$-LAT Collaboration acknowledges support from a number of agencies and institutes for both 
development and the operation of the LAT as well as scientific data analysis. These include NASA and 
DOE in the United States, CEA/Irfu and IN2P3/CNRS in France, ASI and INFN in Italy, MEXT, KEK, and JAXA 
in Japan, and the K.~A.~Wallenberg Foundation, the Swedish Research Council and the National Space Board
in Sweden. Additional support from INAF in Italy and CNES in France for science analysis during the
operations phase is also gratefully acknowledged.~~This study makes use of 43 GHz VLBA data from the
VLBA-BU Blazar Monitoring Program (VLBA-BU-BLAZAR; http://www.bu.edu/blazars/VLBAproject.html). 
This research has made use of data obtained with the Global Millimeter VLBI Array (GMVA), which consists of 
telescopes operated by the MPIfR, IRAM, Onsala, Mets\"ahovi, Yebes and the VLBA. The data were correlated at
the correlator of the MPIfR in Bonn, Germany. The VLBA is an instrument of the National Radio Astronomy 
Observatory, a facility of the National Science Foundation operated under cooperative agreement by Associated
Universities, Inc.


\begin{thebibliography}{28}

\bibitem[{{Abdo} {et~al.}(2009){Abdo}, {Ackermann}, {Ajello}, {Atwood},
  {Axelsson}, {Baldini}, {Ballet}, {Band}, {Barbiellini}, {Bastieri}, \&
  et~al.}]{abdo2009}
{Abdo}, A.~A., {Ackermann}, M., {Ajello}, M., {et~al.} 2009, ApJS, 183, 46

\bibitem[{{Abdo} {et~al.}(2010){Abdo}, {Ackermann}, {Ajello}, {Axelsson},
  {Baldini}, {Ballet}, {Barbiellini}, {Bastieri}, {Baughman}, {Bechtol}, \&
  et~al.}]{abdo2010c}
{Abdo}, A.~A., {Ackermann}, M., {Ajello}, M., {et~al.} 2010, Nature, 463, 919

\bibitem[{{Acero} {et~al.}(2015){Acero}, {Ackermann}, {Ajello}, {Albert},
  {Atwood}, {Axelsson}, {Baldini}, {Ballet}, {Barbiellini}, {Bastieri},
  {Belfiore}, {Bellazzini}, {Bissaldi}, {Blandford}, {Bloom}, {Bogart},
  {Bonino}, {Bottacini}, {Bregeon}, {Britto}, {Bruel}, {Buehler}, {Burnett},
  {Buson}, {Caliandro}, {Cameron}, {Caputo}, {Caragiulo}, {Caraveo},
  {Casandjian}, {Cavazzuti}, {Charles}, {Chaves}, {Chekhtman}, {Cheung},
  {Chiang}, {Chiaro}, {Ciprini}, {Claus}, {Cohen-Tanugi}, {Cominsky}, {Conrad},
  {Cutini}, {D'Ammando}, {de Angelis}, {DeKlotz}, {de Palma}, {Desiante},
  {Digel}, {Di Venere}, {Drell}, {Dubois}, {Dumora}, {Favuzzi}, {Fegan},
  {Ferrara}, {Finke}, {Franckowiak}, {Fukazawa}, {Funk}, {Fusco}, {Gargano},
  {Gasparrini}, {Giebels}, {Giglietto}, {Giommi}, {Giordano}, {Giroletti},
  {Glanzman}, {Godfrey}, {Grenier}, {Grondin}, {Grove}, {Guillemot}, {Guiriec},
  {Hadasch}, {Harding}, {Hays}, {Hewitt}, {Hill}, {Horan}, {Iafrate}, {Jogler},
  {J{\'o}hannesson}, {Johnson}, {Johnson}, {Johnson}, {Johnson}, {Kamae},
  {Kataoka}, {Katsuta}, {Kuss}, {La Mura}, {Landriu}, {Larsson}, {Latronico},
  {Lemoine-Goumard}, {Li}, {Li}, {Longo}, {Loparco}, {Lott}, {Lovellette},
  {Lubrano}, {Madejski}, {Massaro}, {Mayer}, {Mazziotta}, {McEnery},
  {Michelson}, {Mirabal}, {Mizuno}, {Moiseev}, {Mongelli}, {Monzani},
  {Morselli}, {Moskalenko}, {Murgia}, {Nuss}, {Ohno}, {Ohsugi}, {Omodei},
  {Orienti}, {Orlando}, {Ormes}, {Paneque}, {Panetta}, {Perkins},
  {Pesce-Rollins}, {Piron}, {Pivato}, {Porter}, {Racusin}, {Rando}, {Razzano},
  {Razzaque}, {Reimer}, {Reimer}, {Reposeur}, {Rochester}, {Romani},
  {Salvetti}, {S{\'a}nchez-Conde}, {Saz Parkinson}, {Schulz}, {Siskind},
  {Smith}, {Spada}, {Spandre}, {Spinelli}, {Stephens}, {Strong}, {Suson},
  {Takahashi}, {Takahashi}, {Tanaka}, {Thayer}, {Thayer}, {Thompson},
  {Tibaldo}, {Tibolla}, {Torres}, {Torresi}, {Tosti}, {Troja}, {Van Klaveren},
  {Vianello}, {Winer}, {Wood}, {Wood}, {Zimmer}, \& {Fermi-LAT
  Collaboration}}]{acero2015}
{Acero}, F., {Ackermann}, M., {Ajello}, M., {et~al.} 2015, ApJS, 218, 23

\bibitem[{{Ackermann} {et~al.}(2015){Ackermann}, {Ajello}, {Atwood}, {Baldini},
  {Ballet}, {Barbiellini}, {Bastieri}, {Becerra Gonzalez}, {Bellazzini},
  {Bissaldi}, {Blandford}, {Bloom}, {Bonino}, {Bottacini}, {Brandt}, {Bregeon},
  {Britto}, {Bruel}, {Buehler}, {Buson}, {Caliandro}, {Cameron}, {Caragiulo},
  {Caraveo}, {Carpenter}, {Casandjian}, {Cavazzuti}, {Cecchi}, {Charles},
  {Chekhtman}, {Cheung}, {Chiang}, {Chiaro}, {Ciprini}, {Claus},
  {Cohen-Tanugi}, {Cominsky}, {Conrad}, {Cutini}, {D'Abrusco}, {D'Ammando}, {de
  Angelis}, {Desiante}, {Digel}, {Di Venere}, {Drell}, {Favuzzi}, {Fegan},
  {Ferrara}, {Finke}, {Focke}, {Franckowiak}, {Fuhrmann}, {Fukazawa},
  {Furniss}, {Fusco}, {Gargano}, {Gasparrini}, {Giglietto}, {Giommi},
  {Giordano}, {Giroletti}, {Glanzman}, {Godfrey}, {Grenier}, {Grove},
  {Guiriec}, {Hewitt}, {Hill}, {Horan}, {Itoh}, {J{\'o}hannesson}, {Johnson},
  {Johnson}, {Kataoka}, {Kawano}, {Krauss}, {Kuss}, {La Mura}, {Larsson},
  {Latronico}, {Leto}, {Li}, {Li}, {Longo}, {Loparco}, {Lott}, {Lovellette},
  {Lubrano}, {Madejski}, {Mayer}, {Mazziotta}, {McEnery}, {Michelson},
  {Mizuno}, {Moiseev}, {Monzani}, {Morselli}, {Moskalenko}, {Murgia}, {Nuss},
  {Ohno}, {Ohsugi}, {Ojha}, {Omodei}, {Orienti}, {Orlando}, {Paggi}, {Paneque},
  {Perkins}, {Pesce-Rollins}, {Piron}, {Pivato}, {Porter}, {Rain{\`o}},
  {Rando}, {Razzano}, {Razzaque}, {Reimer}, {Reimer}, {Romani}, {Salvetti},
  {Schaal}, {Schinzel}, {Schulz}, {Sgr{\`o}}, {Siskind}, {Sokolovsky}, {Spada},
  {Spandre}, {Spinelli}, {Stawarz}, {Suson}, {Takahashi}, {Takahashi},
  {Tanaka}, {Thayer}, {Thayer}, {Tibaldo}, {Torres}, {Torresi}, {Tosti},
  {Troja}, {Uchiyama}, {Vianello}, {Winer}, {Wood}, \&
  {Zimmer}}]{ackermann2015}
{Ackermann}, M., {Ajello}, M., {Atwood}, W.~B., {et~al.} 2015, ApJ, 810, 14

\bibitem[{{Agudo} {et~al.}(2011){Agudo}, {Jorstad}, {Marscher}, {Larionov},
  {G{\'o}mez}, {L{\"a}hteenm{\"a}ki}, {Gurwell}, {Smith}, {Wiesemeyer}, {Thum},
  {Heidt}, {Blinov}, {D'Arcangelo}, {Hagen-Thorn}, {Morozova}, {Nieppola},
  {Roca-Sogorb}, {Schmidt}, {Taylor}, {Tornikoski}, \& {Troitsky}}]{agudo2011}
{Agudo}, I., {Jorstad}, S.~G., {Marscher}, A.~P., {et~al.} 2011, ApJL, 726,
  L13

\bibitem[{{Ajello} {et~al.}(2015){Ajello}, {Gasparrini}, {S{\'a}nchez-Conde},
  {Zaharijas}, {Gustafsson}, {Cohen-Tanugi}, {Dermer}, {Inoue}, {Hartmann},
  {Ackermann}, {Bechtol}, {Franckowiak}, {Reimer}, {Romani}, \&
  {Strong}}]{ajello2015}
{Ajello}, M., {Gasparrini}, D., {S{\'a}nchez-Conde}, M., {et~al.} 2015, ApJL,
  800, L27

\bibitem[{{Blandford} {et~al.}(2015){Blandford}, {East}, {Nalewajko}, {Yuan},
  \& {Zrake}}]{blanford2015}
{Blandford}, R., {East}, W., {Nalewajko}, K., {Yuan}, Y., \& {Zrake}, J. 2015,
  ArXiv:1511.07515

\bibitem[{{Blandford}(1973)}]{blanford1973}
{Blandford}, R.~D. 1973, A\&A, 26, 161

\bibitem[{{B{\"o}ttcher} {et~al.}(2012){B{\"o}ttcher}, {Harris}, \&
  {Krawczynski}}]{bottcher2012}
{B{\"o}ttcher}, M., {Harris}, D.~E., \& {Krawczynski}, H., eds, 2012, {Relativistic Jets
  from Active Galactic Nuclei}. Wiley, Berlin

\bibitem[{{B{\"o}ttcher} {et~al.}(2007){B{\"o}ttcher}, {Basu}, {Joshi},
  {Villata}, {Arai}, {Aryan}, {Asfandiyarov}, {Bach}, {Bachev}, {Berduygin},
  {Blaek}, {Buemi}, {Castro-Tirado}, {De Ugarte Postigo}, {Frasca}, {Fuhrmann},
  {Hagen-Thorn}, {Henson}, {Hovatta}, {Hudec}, {Ibrahimov}, {Ishii},
  {Ivanidze}, {Jel{\'{\i}}nek}, {Kamada}, {Kapanadze}, {Katsuura}, {Kotaka},
  {Kovalev}, {Kovalev}, {Kub{\'a}nek}, {Kurosaki}, {Kurtanidze},
  {L{\"a}hteenm{\"a}ki}, {Lanteri}, {Larionov}, {Larionova}, {Lee}, {Leto},
  {Lindfors}, {Marilli}, {Marshall}, {Miller}, {Mingaliev}, {Mirabal},
  {Mizoguchi}, {Nakamura}, {Nieppola}, {Nikolashvili}, {Nilsson}, {Nishiyama},
  {Ohlert}, {Osterman}, {Pak}, {Pasanen}, {Peters}, {Pursimo}, {Raiteri},
  {Robertson}, {Robertson}, {Ryle}, {Sadakane}, {Sadun}, {Sigua}, {Sohn},
  {Strigachev}, {Sumitomo}, {Takalo}, {Tamesue}, {Tanaka}, {Thorstensen},
  {Tosti}, {Trigilio}, {Umana}, {Vennes}, {Vitek}, {Volvach}, {Webb},
  {Yamanaka}, \& {Yim}}]{bottcher2007}
{B{\"o}ttcher}, M., {Basu}, S., {Joshi}, M., {et~al.} 2007, ApJ, 670, 968

\bibitem[{{B{\"o}ttcher} {et~al.}(2013){B{\"o}ttcher}, {Reimer}, {Sweeney}, \&
  {Prakash}}]{bottcher2013}
{B{\"o}ttcher}, M., {Reimer}, A., {Sweeney}, K., \& {Prakash}, A. 2013, ApJ,
  768, 54

\bibitem[{{Edelson} \& {Krolik}(1988)}]{edelson1988}
{Edelson}, R.~A. \& {Krolik}, J.~H. 1988, ApJ, 333, 646

\bibitem[{{Finke} \& {Dermer}(2010)}]{finke2010}
{Finke}, J.~D. \& {Dermer}, C.~D. 2010, ApJL, 714, L303

\bibitem[{{Fuhrmann} {et~al.}(2014){Fuhrmann}, {Larsson}, {Chiang},
  {Angelakis}, {Zensus}, {Nestoras}, {Krichbaum}, {Ungerechts}, {Sievers},
  {Pavlidou}, {Readhead}, {Max-Moerbeck}, \& {Pearson}}]{fuhrmann2014}
{Fuhrmann}, L., {Larsson}, S., {Chiang}, J., {et~al.} 2014, MNRAS, 441, 1899

\bibitem[{{Jorstad} {et~al.}(2013){Jorstad}, {Marscher}, {Larionov},
  {G{\'o}mez}, {Agudo}, {Angelakis}, {Casadio}, {Gurwell}, {Hovatta}, {Joshi},
  {Fuhrmann}, {Karamanavis}, {L{\"a}hteenm{\"a}ki}, {Molina}, {Morozova},
  {Myserlis}, {Troitsky}, {Ungerechts}, \& {Zensus}}]{jorstad2013}
{Jorstad}, S., {Marscher}, A., {Larionov}, V., {et~al.} 2013, in European
  Physical Journal Web of Conferences, Vol.~61, European Physical Journal Web
  of Conferences, 04003

\bibitem[{{Jorstad} {et~al.}(2010){Jorstad}, {Marscher}, {Larionov}, {Agudo},
  {Smith}, {Gurwell}, {L{\"a}hteenm{\"a}ki}, {Tornikoski}, {Markowitz},
  {Arkharov}, {Blinov}, {Chatterjee}, {D'Arcangelo}, {Falcone}, {G{\'o}mez},
  {Hagen-Thorn}, {Jordan}, {Kimeridze}, {Konstantinova}, {Kopatskaya},
  {Kurtanidze}, {Larionova}, {Larionova}, {McHardy}, {Melnichuk},
  {Roca-Sogorb}, {Schmidt}, {Skiff}, {Taylor}, {Thum}, {Troitsky}, \&
  {Wiesemeyer}}]{jorstad2010}
{Jorstad}, S.~G., {Marscher}, A.~P., {Larionov}, V.~M., {et~al.} 2010, ApJ,
  715, 362

\bibitem[{{Jorstad} {et~al.}(2001){Jorstad}, {Marscher}, {Mattox}, {Aller},
  {Aller}, {Wehrle}, \& {Bloom}}]{jorstad2001}
{Jorstad}, S.~G., {Marscher}, A.~P., {Mattox}, J.~R., {et~al.} 2001, ApJ, 556,
  738

\bibitem[{{Kagan} {et~al.}(2015){Kagan}, {Sironi}, {Cerutti}, \&
  {Giannios}}]{kagan2014}
{Kagan}, D., {Sironi}, L., {Cerutti}, B., \& {Giannios}, D. 2015, Space Science Reviews, 191,
  545

\bibitem[{{Marscher}(2014)}]{marscher2014}
{Marscher}, A.~P. 2014, ApJ, 780, 87

\bibitem[{{Marscher} {et~al.}(2008){Marscher}, {Jorstad}, {D'Arcangelo},
  {Smith}, {Williams}, {Larionov}, {Oh}, {Olmstead}, {Aller}, {Aller},
  {McHardy}, {L{\"a}hteenm{\"a}ki}, {Tornikoski}, {Valtaoja}, {Hagen-Thorn},
  {Kopatskaya}, {Gear}, {Tosti}, {Kurtanidze}, {Nikolashvili}, {Sigua},
  {Miller}, \& {Ryle}}]{marscher2008}
{Marscher}, A.~P., {Jorstad}, S.~G., {D'Arcangelo}, F.~D., {et~al.} 2008, Nature,
  452, 966

\bibitem[{{Marscher} {et~al.}(2010){Marscher}, {Jorstad}, {Larionov}, {Aller},
  {Aller}, {L{\"a}hteenm{\"a}ki}, {Agudo}, {Smith}, {Gurwell}, {Hagen-Thorn},
  {Konstantinova}, {Larionova}, {Larionova}, {Melnichuk}, {Blinov},
  {Kopatskaya}, {Troitsky}, {Tornikoski}, {Hovatta}, {Schmidt}, {D'Arcangelo},
  {Bhattarai}, {Taylor}, {Olmstead}, {Manne-Nicholas}, {Roca-Sogorb},
  {G{\'o}mez}, {McHardy}, {Kurtanidze}, {Nikolashvili}, {Kimeridze}, \&
  {Sigua}}]{marscher2010}
{Marscher}, A.~P., {Jorstad}, S.~G., {Larionov}, V.~M., {et~al.} 2010, ApJL,
  710, L126

\bibitem[{{Poutanen} \& {Stern}(2010)}]{poutanen2010}
{Poutanen}, J. \& {Stern}, B. 2010, ApJL, 717, L118

\bibitem[{{Rani} {et~al.}(2013{\natexlab{a}}){Rani}, {Krichbaum}, {Fuhrmann},
  {B{\"o}ttcher}, {Lott}, {Aller}, {Aller}, {Angelakis}, {Bach}, {Bastieri},
  {Falcone}, {Fukazawa}, {Gabanyi}, {Gupta}, {Gurwell}, {Itoh}, {Kawabata},
  {Krips}, {L{\"a}hteenm{\"a}ki}, {Liu}, {Marchili}, {Max-Moerbeck},
  {Nestoras}, {Nieppola}, {Quintana-Lacaci}, {Readhead}, {Richards}, {Sasada},
  {Sievers}, {Sokolovsky}, {Stroh}, {Tammi}, {Tornikoski}, {Uemura},
  {Ungerechts}, {Urano}, \& {Zensus}}]{rani2013a}
{Rani}, B., {Krichbaum}, T.~P., {Fuhrmann}, L., {et~al.} 2013{\natexlab{a}},
  A\&A, 552, A11

\bibitem[{{Rani} {et~al.}(2013{\natexlab{b}}){Rani}, {Krichbaum}, {Lott},
  {Fuhrmann}, \& {Zensus}}]{rani2013b}
{Rani}, B., {Krichbaum}, T.~P., {Lott}, B., {Fuhrmann}, L., \& {Zensus}, J.~A.
  2013{\natexlab{b}}, Advances in Space Research, 51, 2358

\bibitem[{{Rani} {et~al.}(2015){Rani}, {Krichbaum}, {Marscher}, {Hodgson},
  {Fuhrmann}, {Angelakis}, {Britzen}, \& {Zensus}}]{rani2015}
{Rani}, B., {Krichbaum}, T.~P., {Marscher}, A.~P., {et~al.} 2015, A\&A, 578,
  A123

\bibitem[{{Rani} {et~al.}(2014){Rani}, {Krichbaum}, {Marscher}, {Jorstad},
  {Hodgson}, {Fuhrmann}, \& {Zensus}}]{rani2014}
{Rani}, B., {Krichbaum}, T.~P., {Marscher}, A.~P., {et~al.} 2014, A\&A, 571, L2

\bibitem[{{Rani} {et~al.}(2013{\natexlab{c}}){Rani}, {Lott}, {Krichbaum},
  {Fuhrmann}, \& {Zensus}}]{rani2013_3c273}
{Rani}, B., {Lott}, B., {Krichbaum}, T.~P., {Fuhrmann}, L., \& {Zensus}, J.~A.
  2013{\natexlab{c}}, A\&A, 557, A71

\bibitem[{{Schinzel} {et~al.}(2012){Schinzel}, {Lobanov}, {Taylor}, {Jorstad},
  {Marscher}, \& {Zensus}}]{schinzel2012}
{Schinzel}, F.~K., {Lobanov}, A.~P., {Taylor}, G.~B., {et~al.} 2012, A\&A, 537,
  A70

\end{thebibliography}


\end{document}